# A Physics-based Analytical Model for Perovskite Solar Cells

Xingshu Sun[1*], Reza Asadpour[1], Wanyi Nie[2], Aditya D. Mohite[2], Muhammad A. Alam[1]

[1]School of Electrical and Computer Engineering, Purdue University, West Lafayette, USA
[2] Materials Physics and Application Division, Los Alamos National Laboratory, Los Alamos, USA.

*Abstract* — **Perovskites are promising next-generation absorber materials for low-cost and high-efficiency solar cells. Although perovskite cells are configured similar to the classical solar cells, their operation is unique and requires development of a new physical model for characterization, optimization of the cells, and prediction of the panel performance. In this paper, we develop such a physics-based analytical model to describe the operation of different types of perovskite solar cells, explicitly accounting non-uniform generation, carrier selective transport layers, and voltage-dependent carrier collection. The model would allow experimentalists to characterize key parameters of existing cells, understand performance bottlenecks, and predict performance of perovskite-based solar panel – the obvious next step to the evolution of perovskite solar cell technology.**

## I. INTRODUCTION

Solar cells have emerged as an important source of renewable energy; further reduction in cost will ensure a broader and accelerated adoption. Recently, organic-inorganic hybrid perovskites, such as $CH_3NH_3PbI_3$, have shown great promise as new absorber materials for low-cost, highly efficient solar cells [1]–[3]. Despite a growing literature on the topic, most of theoretical work to date has been empirical or fully numerical [4]–[8]. The detailed numerical models provide deep insights into the operation of the cells and its fundamental performance bottlenecks; but are generally unsuitable for fast characterization, screening, and/or prediction of panel performance. Indeed, the field still lacks an intuitively simple physics-based analytical model that can interpret the essence of device operation with relatively few parameters, which can be used to characterize, screen, and optimize perovskite-based solar cells, provide preliminary results for more sophisticated device simulation, and allow panel-level simulation for perovskites. This state-of-art reflects the fact that despite a superficial similarity with p-n [9]–[11] or p-i-n [12]–[14] solar cells, the structure, self-doping, and charge collection in perovskite cells are unique, and cannot described by traditional approaches [15], [16].

This work is supported by the U.S. Department of Energy under DOE Cooperative Agreement no. DE-EE0004946 ("PVMI Bay Area PV Consortium"), the National Science Foundation through the NCN-NEEDS program, contract 1227020-EEC, and by the Semiconductor Research Corporation.

The authors are with the Department of Electrical and Computer Engineering, Purdue University, West Lafayette, IN 47907 USA (e-mail: sun106@purdue.edu; rasadpou@purdue.edu; alam@purdue.edu), the materials physics and application division, Los Alamos National Laboratory (wanyi@lanl.gov; amohite@lanl.gov).

In this paper, we present a new physics-based analytical model that captures the essential features of perovskite cells, namely, position-dependent photo-generation, the role of carrier transport layers, *e.g.*, $TiO_2$ and Spiro-OMeTAD, in blocking charge loss at wrong contacts, voltage-dependent carrier collection that depends on the degree of self-doping of the absorber layer, etc. The model is systematically validated against the four classes of perovskite solar cells reported in the literature. We demonstrate how the model can be used to obtain physical parameters of a cell and how the efficiency can be improved. Our model can be easily converted into a physics-based equivalent circuit that is essential for accurate and complex large-scale network simulation to evaluate and optimize perovskite-based solar modules and panels [13], [17]–[20].

## II. MODEL DEVELOPMENT AND VALIDATION

A typical cell consists of a perovskite absorber layer (300 ~ 500 nm), a hole transport layer (p-type), an electron transport layer (n-type), and front and back contacts, arranged in various configurations. The traditional structure in Fig. 1 (a, b) has PEDOT: PSS and PCBM as the front hole transport layer and the back electron transport layer, respectively; in the inverted structure, however, $TiO_2$ is the front electron transport layer and Spiro-OMeTAD is the back hole transport layer, as in Fig. 1 (c, d). Moreover, for both the traditional and inverted configurations, it has been argued that the absorber layer in high-efficiency cells is essentially intrinsic [21], see Fig. 1 (a,c); the mode of operation changes and the efficiency is reduced for cells with significant p-type self-doping [22], see Fig. 1 (b,d). Therefore, perovskite solar cells can be grouped into (Type-1) p-i-n, (Type-2) p-p-n, (Type-3) n-i-p, (Type-4) n-p-p cells; the corresponding energy band diagrams are shown in Fig. 1.

It has been suggested that the high dielectric constant of perovskites allows the photogenerated excitons to dissociate immediately into free carriers [23], [24]. The photo-generated electron and holes then drift and diffuse through the absorber and transport layers before being collected by the contacts. Consequently, an analytical model can be developed by solving the steady state electron and hole continuity equations within the absorber, namely,

$$D\frac{\partial^2 n(x)}{\partial x^2} + \mu E(x)\frac{\partial n(x)}{\partial x} + G(x) - R(x) = 0. \quad (1)$$

$$D\frac{\partial^2 p(x)}{\partial x^2} - \mu E(x)\frac{\partial p(x)}{\partial x} + G(x) - R(x) = 0. \quad (2)$$

Here, $n(p)$ is the electron/hole concentration; $D$ and $\mu$ are the diffusion coefficient and mobility, respectively; and $G(x)$ represents the position-dependent photo-generation. The extraordinarily long diffusion length in perovskite [25]–[27] ensure that one can ignore carrier recombination within the absorber layer, i.e., $R(x) = 0$. Finally, $E(x)$ is the position-resolved electric field within the absorber layer.

As shown in Fig. 1, $E(x)$ is a constant (linear potential profile) for type-1 (n-i-p) and type-3 (p-i-n) cells, i.e., the absence of doping or trapped charges ensure that $E(x) = (V_{bi} - V)/t_0$, where $V_{bi}$ is the build-in potential and $t_0$ is the thickness of the intrinsic layer. For type-2 (p-p-n) and type -4 (n-p-p) devices, however, numerical simulation shows that the field essentially linear within the depletion region, i.e., $E(x) = [1 - x/W_d] E_{max}(V)$, where $W_d$ is the depletion width and $|E_{max}(V)| = 2(V_{bi} - V)/W_d(V)$ ; $E(x) = 0$ in the neutral region defined by $x > W_d$. The position-dependent $E(x)$ is reflected in the parabolic potential profiles shown in Fig. 1 (b) and (d). Our extensive numerical simulation [21] shows that the photogenerated carriers do not perturb the electric field significantly, therefore, the following analysis will presume $E(x)$ is independent of photogeneration at 1-sun illumination.

Neglecting any parasitic reflectance from the back surface, we approximate the generated profile in the absorber layer as $G(x) = G_{eff} e^{-x/\lambda_{ave}}$, where $G_{eff}$ and $\lambda_{ave}$(~100 nm) are the material specific constants, averaged over the solar spectrum. Note that the maximum absorption is $G_{max} = \int_0^\infty G_{eff} e^{-x/\lambda_{ave}} dx = G_{eff} \lambda_{ave}$.

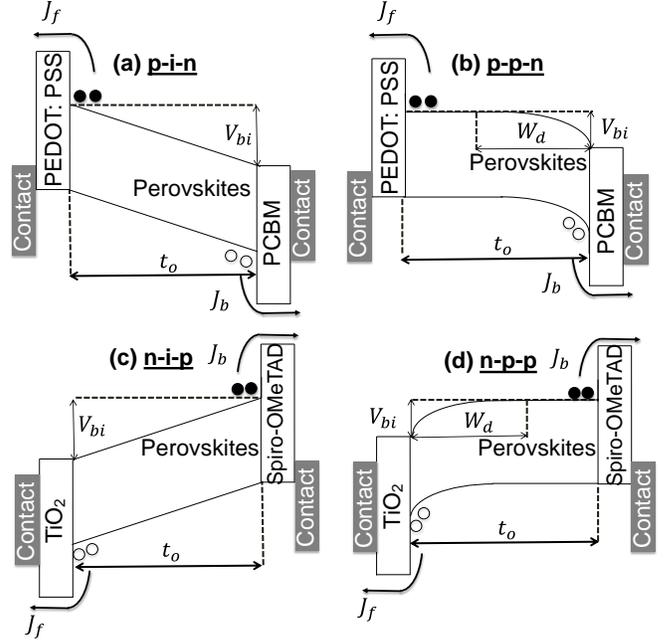

Fig. 1. The energy diagram of perovskite solar cells in traditional structure (PEDOT: PSS/ Perovskite/PCBM): (a) Type-1 (p-i-n) and (b) Type-2 (p-p-n) and titania-based inverted cells (TiO2/Perovskite/Spiro-OMeTAD): (c) Type-3 (n-i-p) and (d) Type-4 (n-p-p).

Finally, electron and hole transport layers are considered perfect conductors for the majority carriers; while they act as imperfect blocking layers for the minority carriers, characterized by the effective surface recombination velocity $|J_{f(b)}| = q s_{f(b)} \Delta n(p)$. The $\Delta n(p)$ is the excess minority carrier concentration, and the $s_{f(b)}$ is the effective surface recombination velocity at the front (back) transport layer, accounting for three recombination processes: 1) carriers escape from the wrong contact; 2) recombination due to the interface defects; 3) recombination within the bulk of the transport layer.

TABLE I. Model parameters of Eqs. (5)-(7) expressed in terms of the physical parameters of the cell. Here, $(V' = q(V - V_{bi})/kT$; $\beta_{f(b)} = D/(t_o \times s_{f(b)})$; $m = t_o/\lambda_{ave}$; $n = W_d(0\,V)/t_o$; $\Delta = 1 - n\sqrt{(V_{bi} - V)/V_{bi}}$. The meaning of the parameters has been discussed in the text.

| Variables | p-i-n / n-i-p | p-p-n | n-p-p |
|---|---|---|---|
| $1/\alpha_f$ | $\dfrac{e^{V'} - 1}{V'} + \beta_f$ | $\Delta + \beta_f$ $(V \leq V_{bi})$ | $\Delta \times e^{V'} + \beta_f$ $(V \leq V_{bi})$ |
| | | $\dfrac{e^{V'} - 1}{V'} + \beta_f$ $(V > V_{bi})$ | |
| $1/\alpha_b$ | $\dfrac{e^{V'} - 1}{V'} + \beta_b$ | $\Delta \times e^{V'} + \beta_b$ $(V \leq V_{bi})$ | $\Delta + \beta_b$ $(V \leq V_{bi})$ |
| | | $\dfrac{e^{V'} - 1}{V'} + \beta_b$ $(V > V_{bi})$ | |
| $A$ | $\alpha_f \times \left(\dfrac{1 - e^{V' - m}}{V' - m} - \beta_f\right)$ | $\alpha_f \times \left(\dfrac{1}{m}(e^{-m \times \Delta} - 1) - \beta_f\right) (V \leq V_{bi})$ | $\alpha_f \times \left(\dfrac{e^{V'}}{m}(e^{-m} - e^{m \times (\Delta - 1)}) - \beta_f\right) (V \leq V_{bi})$ |
| | | $a_f \times \left(\dfrac{(1 - e^{V' - m})}{V' - m} - \beta_f\right) (V > V_{bi})$ | |
| $B$ | $a_b \times \left(\dfrac{1 - e^{V' + m}}{V' + m} - \beta_b\right)$ | $\alpha_b \times \left(\dfrac{e^{V'}}{m}(e^{-m \times (\Delta - 1)} - e^m) - \beta_b\right) (V \leq V_{bi})$ | $\alpha_b \times \left(\dfrac{1}{m}(1 - e^{m \times \Delta}) - \beta_b\right) (V \leq V_{bi})$ |
| | | $a_b \times \left(\dfrac{(1 - e^{V' + m})}{V' + m} - \beta_b\right) (V > V_{bi})$ | |

Remarkably, Eqs. (1) - (2) can be solved analytically to derive the complete current-voltage characteristics of the four types of perovskite cells, as follows

$$J_{dark} = (\alpha_f \times J_{f0} + \alpha_b \times J_{b0})\left(e^{\frac{qV}{kT}} - 1\right), \quad (3)$$

$$J_{photo} = qG_{max}(A - Be^{-m}), \quad (4)$$

$$J_{light} = J_{dark} + J_{photo}. \quad (5)$$

The parameters of the model, namely, $\alpha_{f(b)}$, $\beta_{f(b)}$, $A(B)$, $m$, $n$, and $\Delta$ are functions of the following physical parameters of the cell (see Table I): $t_0$ is the thickness of the absorber layer; $J_{f0(b0)}$ is the dark diode current recombining at the front/back transport layer; $V_{bi}$ is the built in potential across the absorber layer; $D$ is the diffusion coefficient; $s_{f(b)}$ is the effective surface recombination velocity at the front/back interface; $W_d(0\,V)$ is the equilibrium depletion width for self-doped devices; and $G_{max}$ is the maximum absorption.

Among these parameters, $G_{max}$ is obtained by integrating the position-dependent photon absorption calculated by the transfer matrix method [28] (here $qG_{max} = 23$ mA/cm$^2$); $D \approx 0.05$ cm$^2$s$^{-1}$ is known for the material system for both electron and hole [26]; $V_{bi}$ can be estimated either by using the capacitance-voltage characteristics [22] or by using the crossover voltage of the dark and light IV [29]. The effective surface recombination velocities can be fitted using the photogenerated current $J_{photo}(G,V) = J_{light}(G,V) - J_{dark}(V)$ [30]. Finally, we can obtain the dark diode current $J_{f0/b0}$ by fitting the dark current.

In order to validate the model, we fit both dark and light IV characteristics for four different perovskite cells using the model as shown in Fig. 2. See the appendix for the details of the fitting algorithm. Samples #1 (15.7 %) and #2 (11.1 %) are solution-based PCBM based architecture (Type-1 and Type-2) [21], whereas samples #3 (15.4 %) and #4 (8.6 %) are titania-based inverted architecture (Type-3 and Type-4) fabricated by vapor deposition and solution process, respectively [31]. The fitting parameters obtained for the four samples are summarized in Table II. Remarkably, the analytical model not only reproduces the key features of the I-V characteristics of very different cell geometries, but also captures very well the known physical parameters of the cell (e.g. thickness of the absorber). Indeed, the error in the power output due to imperfect fitting is less than 0.1% (absolute) for samples 1-3, and ~0.5% (absolute) for sample 4.

### III. RESULTS AND DISCUSSION

Fig. 2(b,d) shows that the light IV of the self-doped devices has a steep decrease (~ 0 V – 0.5 V) in photocurrent much before the maximum power point (MPP). Indeed, this characteristic feature can be correlated to self-doping effects arising from the defects or impurities introduced during the manufacture of the cell. Our model interprets this linear decrease in photocurrent of type-2 and type-4 cells to the well-known voltage-dependent reduction of $W_d(V)$ (also the charge collection region) of a PN junction. Without a physics-based model, this feature can be easily mistaken as a parasitic resistance. The self-doped devices also have an inferior $V_{bi}$ and greater $J_{f0(b0)}$ that leads to a lower $V_{OC}$, compared to the intrinsic cells with the same configuration, see Table II. Hence, the main factor that limits the performance of samples #2 and #4 is the reduction of charge collection efficiency due to self-doping effect.

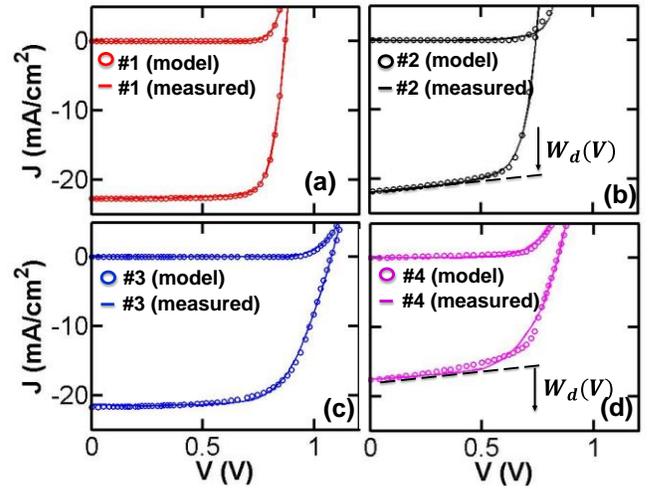

Fig. 2. (a) Samples #1 (Type-1 (p-i-n), Efficiency = 15.7%, $J_{SC}$ = 22.7 mA/cm$^2$, $V_{OC}$ = 0.85 V, FF = 81%). (b) Samples #2 (Type-2 (p-p-n), Efficiency = 11.1%, $J_{SC}$ = 21.9 mA/cm$^2$, $V_{OC}$ = 0.75 V, FF = 64%). (c) Samples #3 (Type-3 (n-i-p), Efficiency = 15.4%, $J_{SC}$ = 21.5 mA/cm$^2$, $V_{OC}$ = 1.07 V, FF = 67%). (d) Samples #4 (Type-4 (n-p-p), Efficiency = 8.6%, $J_{SC}$ = 17.6 mA/cm$^2$, $V_{OC}$ = 0.84 V, FF = 58%). Note that i) $G_{max} = 23$ mA/cm$^2$ is used. ii) Negligible parasitic resistors ($R_{series}$ and $R_{shunt}$) except in samples #4.

TABLE II. Extracted physical parameters of samples #1 (Fig 2 (a)), #2 (Fig 2 (b)), #3 (Fig 2 (c)), and #4 (Fig 2 (d)).

| Sample | #1 | #2 | #3 | #4 |
|---|---|---|---|---|
| Type | p-i-n | p-p-n | n-i-p | n-p-p |
| $t_o$ (nm) | 450 | 400 | 310 | 147 |
| $J_{f0}$ (mA/cm$^2$) | $2.7 \times 10^{-13}$ | $4 \times 10^{-12}$ | $1.6 \times 10^{-17}$ | $6 \times 10^{-15}$ |
| $J_{b0}$ (mA/cm$^2$) | $4 \times 10^{-13}$ | $5 \times 10^{-13}$ | $4.8 \times 10^{-17}$ | $4.1 \times 10^{-13}$ |
| $V_{bi}$ (V) | 0.78 | 0.67 | 1 | 0.75 |
| $s_f$ (cm/s) | $2 \times 10^2$ | $5 \times 10^2$ | $1 \times 10^4$ | 13.1 |
| $s_b$ (cm/s) | 19.2 | $8.6 \times 10^2$ | 5.4 | $\infty$ |
| $W_{depletion}(0\,V)$ (nm) | / | 300 | / | 146 |

While examining the intrinsic samples #1 and #3, we note that #1 has the highest fill-factor (*FF*), but its $V_{OC}$ is 0.3V smaller than that of #3. The reduction in $V_{oc}$ can be explained by lower $V_{bi}$ and higher $J_{f0(b0)}$ caused by the combination of band misalignment and lower doping concentration in the transport layers of the perovskite cells with the traditional structure, which is the major performance limitation of #1. Sample #3, on the other hand, has the lower fill-factor, arising from relatively high effective surface recombination velocities at both contacts, indicating insufficient blocking of charge loss to the wrong contact. Even though #1 and #3 have similar efficiencies, our model demonstrates that the fundamental performance limitations are completely different.

Using the model, we can also extract the thicknesses of the four samples, which are in the expected range (~350 nm – 500 nm for #1 and #3, ~ 330 nm for #2) [21], [31]. Among the samples, there is also a strong correlation between the absorber thickness $t_0$ and $J_{SC}$, related to the completeness of the absorption. Moreover, we observe significant shunt resistance ($R_{shunt} = 1\,k\Omega \cdot cm^2$) in sample #4, which agrees with the reports [31] that thin absorber might lead to shunting pinholes. Further, except for sample #4, all devices have relatively poor (high) $s_{front}$, which may be caused by insufficient barrier between PEDOT:PSS and perovskites [21] as well as poor carrier collection in TiO$_2$ [32]–[34].

Once we extract the physical parameters associated with high-efficiency samples (#1 and #3) with essentially intrinsic absorbers, it is natural to ask if the efficiency could be improved further, and if so, what factors would be most important. The physics-based compact model allows us to explore the phase-space of efficiency as a function of various parameters, as follows.

For example, while keeping all other parameter equal to the values extracted in Table II, one can explore the importance of absorber thickness on cell efficiency, see Fig. 3. Our model shows that both samples are close to their optimal thickness, though there is incomplete absorption ($J_{SC} < qG_{max}$). Thinner absorber cannot absorb light completely, while thicker absorber suppresses charge collection and degrades the fill factor. This is because the competition between the surface recombination and the electric field determines the carrier collection efficiency near the interface, and electric field $E = (V_{bi} - V)/t_o$ decreases with the thickness. To summarize, for the samples considered, thickness optimization would not improve performance.

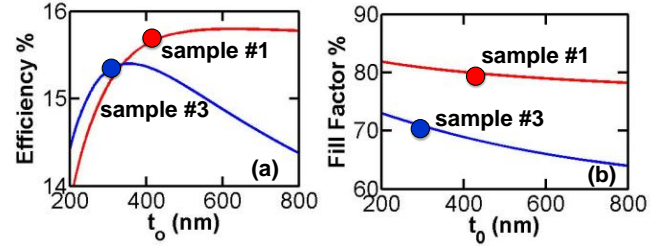

Fig. 3. (a) Efficiency vs. absorber thickness for samples #1 and #3. (b) Fill factor vs. absorber thickness for samples #1 and #3.

Similarly, we can investigate the effects of the front/back surface recombination velocities on device efficiencies, with all other parameters kept fixed to those in Table II. The deduced surface recombination velocities for samples #1 and #3 are listed in Table II as well as labeled as black dots in Fig. 4. The results suggests that, in principle, improving the front surface recombination velocities by two orders of magnitude can boost the efficiency by ~ 3% and even ~5% for samples #1 and #3, respectively. Any potential improvement in the back selective blocking layer, however, offers very little gain, since most of the photo-generation occurs close to the front contact. Hence, engineering the front transport layer would be essential in further improvement of cell efficiencies.

But even with the optimal surface recombination velocities, we are still not close to the thermodynamic limit (~ 30%), see Fig. 4. Towards this goal, one must improve the $J_{SC}$, *FF*, and $V_{OC}$ (thermodynamic limit: $J_{SC}$ ~ 26 mA/cm$^2$, *FF* ~90%, $V_{OC}$ ~ 1.3 V [35]). One may reduce the parasitic absorption loss in the transport layers, which can increase $G_{max}$ in Eq. (4), to improve the $J_{SC}$; one may still improve the *FF* by increasing the charge diffusion coefficient *D*, since it is mainly the variable $\beta_{f(b)} = D/(t_o \times s_{f(b)})$ that determines the *FF*; one may also increase the built-in potential $V_{bi}$, through adjusting the band alignment at the interface as well as increasing the doping of the transport layers, to improve the $V_{OC}$.

We conclude this section with a discussion regarding hysteresis of the J-V characteristics, which can be an important concern for the inverted structure shown in Fig. 1 (c, d)). The phenomenon arises primarily from by trapping/detrapping of defects within the oxide or at the oxide/perovskite interface [32], [33]. Reassuringly, recent results show that process-improvements, such as Li-treatment of TiO$_2$, can suppress/eliminate hysteresis, see [36]. Moreover, cells with the traditional structures (oxide-free, as in Fig. 1 (a, b)) show very little hysteresis [21], [37]. Given the fact that hysteresis effects will be eventually minimized once perovskites are mature enough for integration in modules, the compact model proposed in this paper does not account for the effect of hysteresis explicitly.

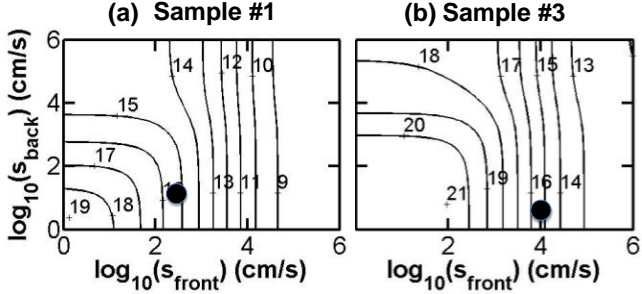

Fig. 4. (a) Contour plot of the front/back surface recombination velocities vs. efficiency for sample #1. (b) Contour plot of the front/back surface recombination velocities vs. efficiency for sample #3.

## V. CONCLUSIONS

We have derived an analytical model that describes both dark and light current-voltage characteristics for four different types [p-i-n/p-p-n and n-i-p/n-p-p] of perovskite solar cells. An important contribution of the model is that, along with other measurement techniques, it provides a simple and complementary approach to characterize, optimize, and screen fabricated cells. Physical parameters that cannot be directly measured, such as $V_{bi}$ of a p-i-n device, can also be deduced using the model.

Apart from determining the parameters of an existing cell and suggesting opportunities for further improvement, an analytical compact model serves another fundamental need, namely, the ability to predict the ultimate performance of the panel composed of individual perovskite cells. Panel efficiency is ultimately dictated by process variation reflected in various parameters (as in Table II) as well as statistical distribution of shunt and series resistances [13], [38]. Indeed, recent studies [39], [40] show large efficiency gap between perovskite-based solar cells and modules – an equivalent circuit based on the physics-based analytical model developed in this paper will be able to trace the cell-module efficiency gap to statistical distribution of one or more cell parameters and suggest opportunities for improvement. Closing this cell-to-module gap is the obvious next step and an essential pre-requisite for eventual commercial viability of the perovskite solar cells.

## APPENDIX

The parameters of the compact model are extracted by fitting the equations to experimental data. The fitting algorithm has two parts: 1) Model choice 2) Iterative fitting.

Before one fits the data, the structure of the cell must be known (e.g., PEDOT: PSS/ Perovskite/PCBM or TiO$_2$/Perovskite/Spiro-OMeTAD) and whether the absorber is self-doped or not. Ideally, the capacitance-voltage measurement provides the doping profile; as an alternative, we find that the steepness (*dI/dV*) of the light I-V curve at low voltage can also differentiate self-doped and intrinsic cells, see Fig. 2. Specifically, the light IV of the self-doped device (sample #2) shows a steep decrease (~ 0 V – 0.5 V) in photocurrent much before the maximum power point (MPP); an undoped device (sample #1), however, shows flat light IV before MPP . If the parasitic resistance extracted from dark IV is not significant, our model attributes this decrease in photocurrent to voltage-dependent reduction of the depletion region (charge collection) of a doped absorber. Such a feature helps one to choose the correct model for a device.

Estimating the initial guesses and limiting the range of each parameter (from physical considerations) is an important step, since the fitting procedure utilize the iterative fitting function "lsqcurvefit" in MATLAB®, whose results depend on the initial guesses significantly.

The physical parameters we attempt to deduce are: $G_{max}$, $\lambda_{ave}$, $t_o$, $W_d(0\text{ V})$ (self-doped), $D$, $s_f$, $s_b$, $V_{bi}$, $J_{f0}$, and $J_{b0}$. Among these parameters, based on the transfer matrix method, $qG_{max}$ can be obtained by integrating the photon absorption (around 23 mA/cm$^2$) and $\lambda_{ave}$ is around 100 nm; $D \approx 0.05 \text{ cm}^2\text{s}^{-1}$ is known for the material system for both electrons and holes.

Presuming the dark current is illumination-independent, one can calculate photocurrent following

$$J_{photo}(G,V) = J_{light}(G,V) - J_{dark}(V). \quad (A1)$$

400 nm is a sensible initial guess for $t_o$, since the absorber thickness is around 300 nm to 500 nm for perovskite solar cells. Though capacitance measurement can determine $W_d(0\text{ V})$ for a self-doped device, one can make $W_d(0\text{ V}) \approx 300$ nm as an initial guess. It has been shown that $s_f$ is inferior to $s_b$ in most cases due to low insufficient barrier between PEDOT:PSS and perovskites as well as low carrier lifetime in TiO$_2$. Hence, the initial guesses for $s_f$ and $s_b$ could be approximately $10^3$ cm/s and $10^2$ cm/s, respectively. The junction built-in $V_{bi}$ is estimated to be the cross-over voltage of dark and light IV curves. Then one can first use the "lsqcurvefit" function to fit the photocurrent based on the initial guesses.

Since $J_{f0}$ and $J_{b0}$ is on the order of $10^{-13}$ to $10^{-15}$ mA/cm$^2$, one can use zero as the initial guesses. Afterwards, one can use the iterative fitting procedure for the dark current while the parameters extracted from photocurrent are fixed.

Once the parameters are obtained, they must be checked for self-consistency and convergence between light and dark characteristics.

ACKNOWLEDGEMENT

This work is supported by the U.S. Department of Energy under DOE Cooperative Agreement no. DE-EE0004946 ("PVMI Bay Area PV Consortium"), the National Science Foundation through the NCN-NEEDS program, contract 1227020-EEC, and by the Semiconductor Research Corporation. The authors would like to thank Raghu Chavali and Ryyan Khan for helpful discussion and Professor Mark Lundstrom for kind guidance.

**Xingshu Sun** (S'13) received the B.S. degree from Purdue University, West Lafayette, IN, in 2012, where he is currently working toward the Ph.D. degree in electrical and computer engineering. His current research interests include device simulation and compact modeling for photovoltaics and nanoscale transistors.

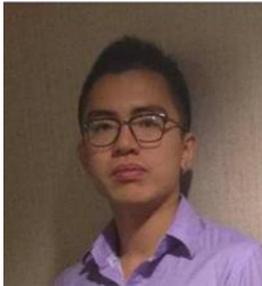

**Reza Asadpour** Reza received his B. Sc. and M. Sc. degrees in electrical engineering from University of Tehran, Tehran, Iran, in 2010 and 2013, respectively. Since 2013, he has been with Prof. Alam's CEED group at Purdue University working on solar cell and their reliability issues towards his Ph.D. degree.

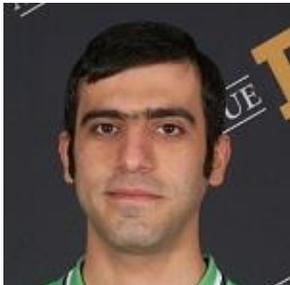

**Wanyi Nie** received the PhD degree in department of physics in Wake Forest University, Winston-Salem, NC. She is currently conducting research as a Postdoc Fellow in Los Alamos National Lab on opto-electronic device research in Material Synthesis and Integrated Device Group, MPA-11. Her research interest is on photovoltaic device physics and interface engineering.

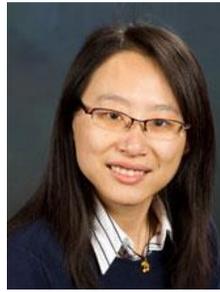

**Aditya D. Mohite** received his B.S. and M.S. degree from Maharaja in Solid State Physics from Maharaja Sayajirao University of Baroda, India and Ph.D. degree from University of Louisville, KY, USA in 2008, all in electrical engineering. He is currently a Staff Scientist with the Materials synthesis and integrated devices group at Los Alamos National Laboratory and directs an optoelectronics group (light to energy team) working on understanding and controlling photo-physical processes in materials for thin film light to energy conversion technologies such as photovoltaics, photo-catalysis etc. He is an expert with correlated techniques such as photocurrent microscopy and optical spectroscopy to investigate the charge and energy transfer and recombination processes in thin-film devices.

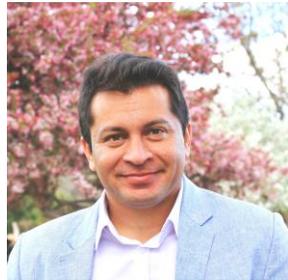

**Muhammad Ashraful Alam** (M'96–SM'01–F'06) is the Jai N. Gupta Professor of Electrical and Computer Engineering where his research and teaching focus on physics, simulation, characterization and technology of classical and emerging electronic devices. From 1995 to 2003, he was with Bell Laboratories, Murray Hill, NJ, where he made important contributions to reliability physics of electronic devices, MOCVD crystal growth, and performance limits of semiconductor lasers. At Purdue, Alam's research has broadened to include flexible electronics, solar cells, and nanobiosensors. He is a fellow of the AAAS, IEEE, and APS and recipient of the 2006 IEEE Kiyo Tomiyasu Award for contributions to device technology.

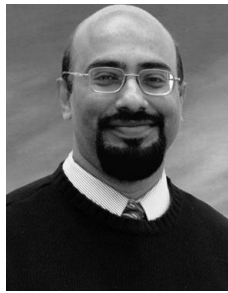

# A Physics-based Analytical Model for Perovskite Solar Cells

Xingshu Sun[1], Reza Asadpour[1], Wanyi Nie[2], Aditya D. Mohite[2] and Muhammad A. Alam[1].

[1]Purdue University School of Electrical and Computer Engineering, West Lafayette, IN, 47907, USA.

[2]Materials Physics and Application Division, Los Alamos National Laboratory, Los Alamos, NM 87545, USA.

# Supplementary Information

## 1. Derivation of Eqs. (5) to (7)

Here we will discuss the analytical derivation of the dark and light IV for perovskite solar cells.

**1.1 Intrinsic absorber: Type 1 (p-i-n) and Type 3 (n-i-p), see Fig. S1.1**

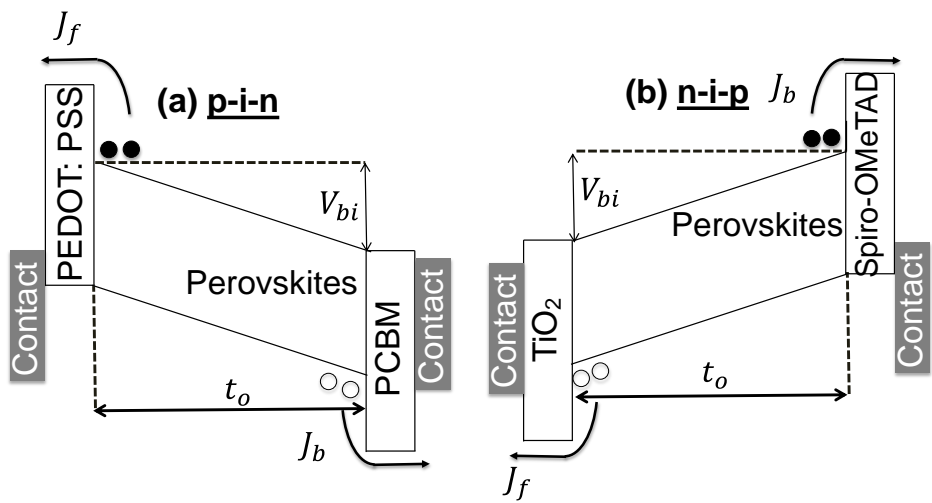

Figure S1.1 (a) The energy diagram of (a) Type 1 (p-i-n) and (b) Type 3 (n-i-p) perovskite cells

We will begin with solving the electron and hole continuity equations given in [1]

$$\frac{\partial n}{\partial t} = \frac{1}{q}\frac{\partial J_n}{\partial x} + G(x) - R(x), \tag{S1.1}$$

$$\frac{\partial p}{\partial t} = -\frac{1}{q}\frac{\partial J_p}{\partial x} + G(x) - R(x), \tag{S1.2}$$

where $n$ and $p$ are the electron and hole concentrations, $G(x)$ and $R(x)$ denote the generation and recombination processes, and $J_n$ and $J_p$ are the electron and hole currents expressed as follows:

$$J_n = q\mu_n nE + qD_n\frac{\partial n}{\partial x}, \tag{S1.3}$$

$$J_p = q\mu_p p E - q D_p \frac{\partial p}{\partial x}. \tag{S1.4}$$

In Eqs. (S1.3) and (S1.4), $E$ is the electric field, $\mu_n$ and $\mu_p$ are the electron and hole motilities, $D_n$ and $D_p$ are the electron and hole diffusion coefficients, respectively.

Assuming that the bulk recombination is negligible (*i.e.*, $R(x) = 0$) [2], Eqs. (S1.1) to (S1.4) reduce to,

$$D_n \frac{\partial^2 n}{\partial x^2} + \mu_n E \frac{\partial n}{\partial x} + G(x) = 0, \tag{S1.5}$$

$$D_p \frac{\partial^2 p}{\partial x^2} - \mu_p E \frac{\partial p}{\partial x} + G(x) = 0. \tag{S1.6}$$

To solve the equations, we first need to calculate $E$ by solving the Poisson equation, and the generation profile, $G(x)$, by solving the Maxwell equations.

The Poisson equation is written as

$$\frac{\partial^2 \phi}{\partial x^2} = -\frac{\rho}{\epsilon}. \tag{S1.7}$$

Assuming that the absorber is intrinsic (so that $\rho = 0$), therefore, $\phi(x) = ax$. Since the voltage drops primarily across the absorber layer, therefore, $\phi(x = 0) = 0 \ and \ \phi(x = t_0) = V_{bi} - V$ in the p-i-n structure. Hence, we can express the electric field as $a = \frac{V_{bi} - V}{t_0} = \frac{d\phi}{dx} = -E,$ so that $E = (V - V_{bi})/t_o$. Recall that $V_{bi}$ is the built-in potential across the absorber that is mainly determined by the doping of the selective transport layers as well as the band alignment at the interface, and $t_o$ is the absorber thickness, see Fig. S1.2 (a).

The generation profile within the absorber can be approximated as $G(x) = G_{eff} e^{-x/\lambda_{ave}}$, provided one neglects back reflectance, see Fig. S1.2 (b). The optical absorption depends on the photon wavelength; $\lambda_{ave}$ should be interpreted as the average optical decay length that accounts for the whole solar spectrum.

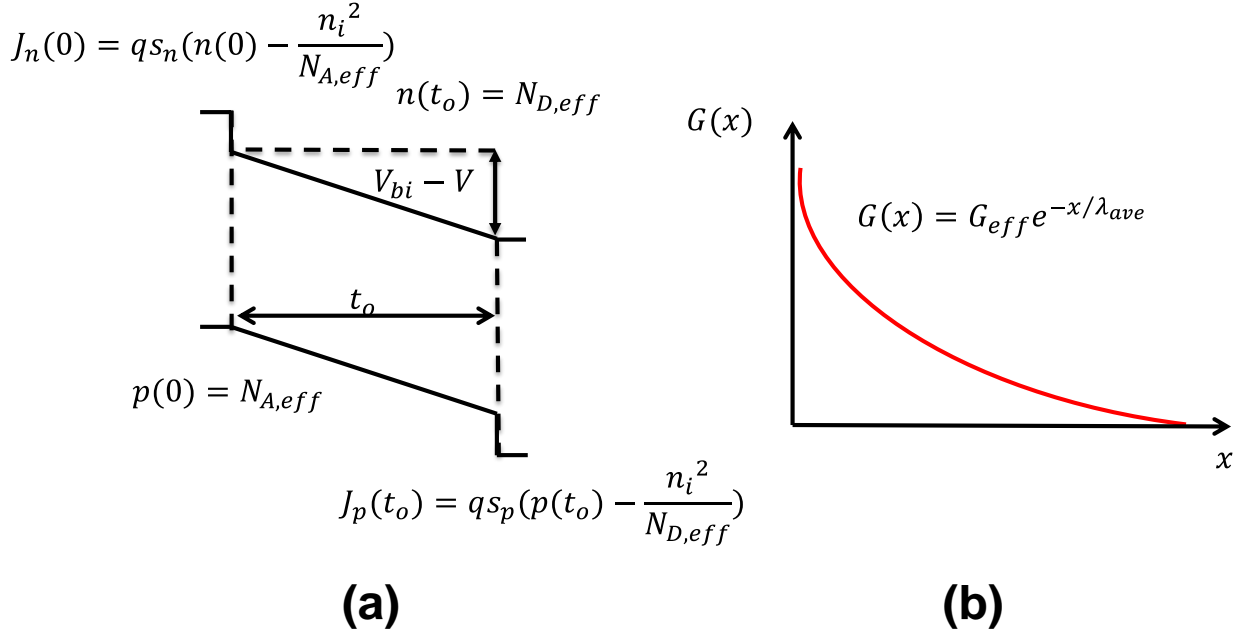

**(a)**                      **(b)**

Figure S1.2 (a) The energy diagram of a p-i-n cell with boundary conditions labeled. (b) The approximated generation profile in the absorber.

After inserting $E$ and $G(x)$ in Eqs. (S1.5) and (S1.6), the general solutions are given by

$$n(x) = A_n e^{-\varepsilon_o x} + \frac{G_n \lambda_{ave}^2 e^{-\frac{x}{\lambda_{ave}}}}{\varepsilon_o \lambda_{ave} - 1} + B_n, \tag{S1.9}$$

$$p(x) = A_p e^{\varepsilon_o x} - \frac{G_p \lambda_{ave}^2 e^{-\frac{x}{\lambda_{ave}}}}{\varepsilon_o \lambda_{ave} + 1} + B_p, \tag{S1.10}$$

where $\varepsilon_o \equiv qE/kT$ is the normalized electric field, $G_n \equiv \frac{G_{eff}}{D_n}$ and $G_p \equiv \frac{G_{eff}}{D_p}$ represent the normalized generation rates, $A_{n(p)}$ and $B_{n(p)}$ are constants to be determined from the boundary conditions.

**In the case of Type 1 (p-i-n)**, the boundary conditions for Eqs. (S1.9) and (S1.10) at $x = 0$ and $x = t_o$ are depicted in Fig. S1.2 (a), where the effective doping concentration $N_{A,eff}$ and $N_{D,eff}$ are the equilibrium hole and electron concentrations at the ends of the i-layer. The concentrations are determined by the doping and the electron affinities of the transport layers, the built-in potential is $V_{bi} = \frac{kT}{q}\log(\frac{N_{A,eff}N_{D,eff}}{n_i^2})$, and $s_n$ and $s_p$ are the minority carrier surface recombination velocities.

Using the boundary conditions, we solve for $B_n$ and $B_p$ as

$$B_n = \frac{N_{D,eff}e^{\varepsilon_o t_o} - \frac{n_i^2}{N_{A,eff}} + \frac{G_n \lambda_{ave}}{\varepsilon_o t_o - 1}(\lambda_{ave} - D_n \frac{\varepsilon_o t_o - 1}{s_n} - \lambda_{ave} e^{\varepsilon_o t - \frac{t_o}{\lambda_{ave}}})}{e^{\varepsilon_o t_o} - 1 + \frac{\varepsilon_o \mu_n kT}{s_n \, q}},$$ (S1.11)

$$B_p = \frac{N_{A,eff}e^{\varepsilon_o t_o} - \frac{n_i^2}{N_{D,eff}} - \frac{G_p \lambda_{ave}}{\varepsilon_o t_o + 1}e^{-\frac{t_o}{\lambda_{ave}}}(\lambda_{ave} - D_p \frac{\varepsilon_o t_o + 1}{s_p} - \lambda_{ave} e^{\varepsilon_o t - \frac{t_o}{\lambda_{ave}}})}{e^{\varepsilon_o t_o} - 1 + \frac{\varepsilon_o \mu_p kT}{s_p \, q}}.$$ (S1.12)

Now utilizing Eqs. (S1.3) and (S1.4), the current density $J = J(0) = J_n(0) + J_p(0)$ can be expressed as $J = qE(\mu_n B_n + \mu_p B_p)$. Substituting Eqs. (S1.11) and (S1.12), we can find the current divided into two parts, a dark diode $J_{dark}$ (independent of generation), and a voltage-dependent photocurrent $J_{photo}$ so that,

$$J_{dark} = \left(\frac{J_{f0}}{\frac{e^{V'}-1}{V'}+\beta_f} + \frac{J_{b0}}{\frac{e^{V'}-1}{V'}+\beta_b}\right)(e^{\frac{qV}{kT}} - 1),$$ (S1.13)

$$J_{photo} = qG_{max}\left(\frac{\frac{(1-e^{V'-m})}{V'-m}-\beta_f}{\frac{e^{V'}-1}{V'}+\beta_f} - \frac{\frac{(1-e^{V'+m})}{V'+m}-\beta_b}{\frac{e^{V'}-1}{V'}+\beta_b}e^{-m}\right),$$ (S1.14)

$$J_{light} = J_{dark} + J_{photo}.$$ (S1.15)

Here, $J_{f0(b0)} = q\frac{n_i^2}{N_{A,eff(D,eff)}}\frac{D_{n(p)}}{t_o}$ is the diode current for electrons and holes recombining at the front or back contact; $\beta_{f(b)} = \frac{D_{n(p)}}{t_o s_{n(p)}}$ depends on the diffusion coefficient and surface recombination velocities; $m = \frac{t_o}{\lambda_{ave}}$ is the ratio of the absorber thickness and the average absorption decay length; $G_{max} = G_{eff}\lambda_{avg}$ is the maximum generation ($G_{max} = \int_o^\infty G_{eff} e^{-x/\lambda_{avg}}dx$); $V'$ represents $q(V - V_{bi})/kT$.

Eqs. (S1.13) to (S1.15) can be further simplified to

$$\alpha_{f(b)} = 1/\left(\frac{e^{V'}-1}{V'} + \beta_{f(b)}\right),$$ (S1.16)

$$A = \alpha_f \times \left(\frac{(1-e^{V'-m})}{V'-m} - \beta_f\right),$$ (S1.17)

$$B = \alpha_b \times \left(\frac{(1-e^{V'+m})}{V'+m} - \beta_b\right).$$ (S1.18)

Consequently,

$$J_{dark} = (\alpha_f \times J_{f0} + \alpha_b \times J_{b0})(e^{\frac{qV}{kT}} - 1),$$ (S1.19)

$$J_{photo} = qG_{max}(A - Be^{-m}). \tag{S1.20}$$

Similarly, one can derive the equations **for Type 3 (n-i-p) perovskite** solar cells with different boundary conditions (i.e., $J_p(0) = qs_p\left(n_i - \frac{n_i^2}{N_{D,eff}}\right)$ and $n(0) = N_{D,eff}$; $J_n(t_o) = qs_n(n_i - \frac{n_i^2}{N_{A,eff}})$ and $p(t_o) = N_{A,eff}$).

**1.2 Self-doped absorber: Type 2 (p-p-n) and Type 4(n-p-p), see Fig. S1.3**

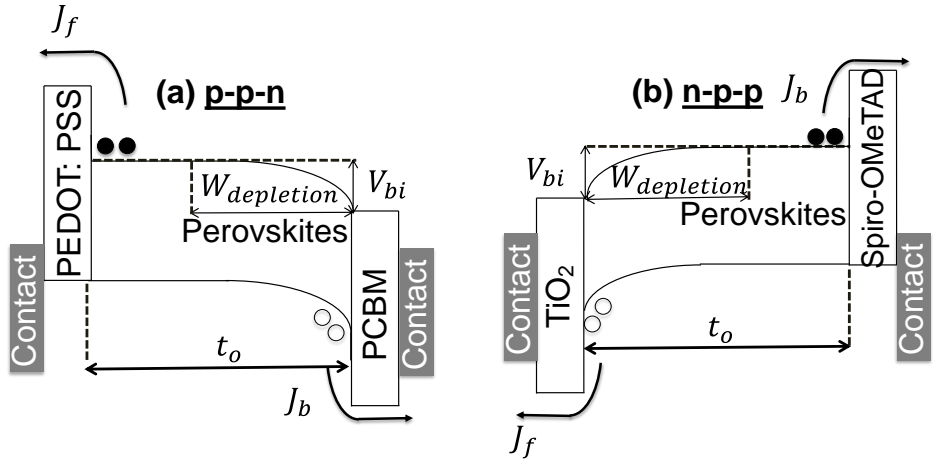

Figure S1.3 (a) The energy diagram of (a) Type 3 (p-p-n) and (b) Type 4 (n-p-p) perovskite cells

Due to the intrinsic defects, perovskite films might be self-doped. Generally, self-doping is more pronounced in low/medium (6 ~ 12%) efficiency devices. Here, we derive a physics-based compact model for both p-p-n and n-p-p structures following a recipe similar to that of p-i-n/n-i-p structures.

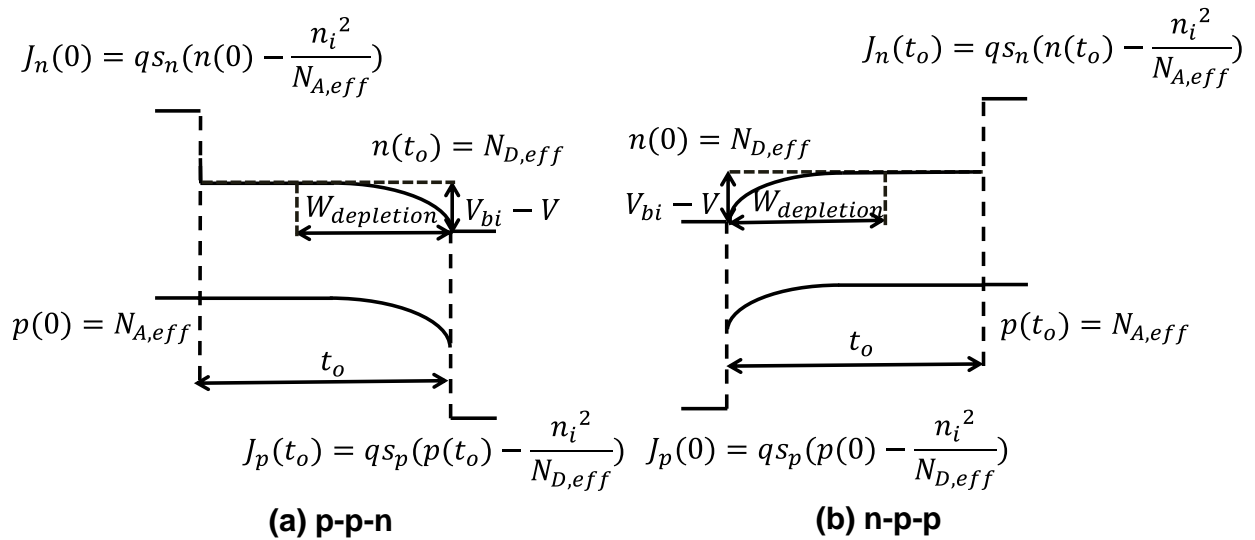

Figure S1.4 The energy diagram of (a) p-p-n and (b) n-p-p perovskite solar cells with boundary conditions labeled.

The energy diagrams of p-p-n and n-p-p structures are shown in Fig. S1.4. The system can be divided into two parts: 1) the depletion region, $W_{delption}(V) = W_{delption}(0\ V)\sqrt{\frac{(V_{bi}-V)}{V_{bi}}}$ ($V < V_{bi}$); 2) the neutral charge region, $t_0 - W_{delption}(V)$. Fig. S1.5 shows the corresponding electric field profiles ($V < V_{bi}$), where the field in the neutral charge regions are zero, while that in the depletion region is presumed linear following $|E_{max}(V)| = \frac{2(V_{bi}-V)}{W_{delp}(V)}$.

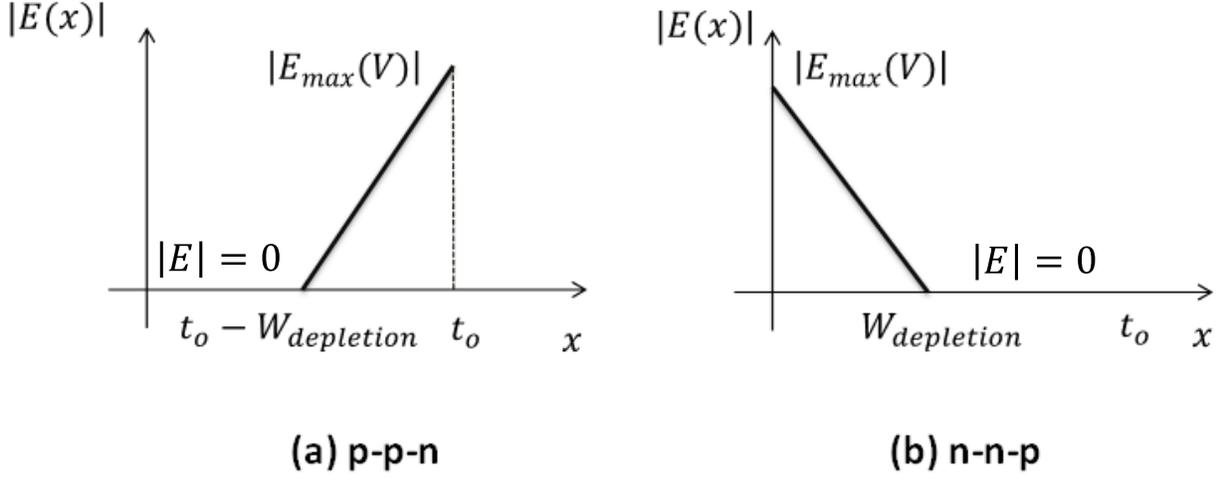

(a) p-p-n  (b) n-n-p

Figure S1.5 Electric field of (a) Type 2 (p-p-n) and (b) Type 4 (n-p-p) perovskite solar cells.

We adopt the same boundary conditions and generation profile as in Section 1.1 to solve Eqs. (S1.5) and (S1.6). Additionally, the charges and the currents must be continuous at the boundary between the depletion and neutral regions, i.e., $J_{n(p)}(l^-) = J_{n(p)}(l^+)$ and $n, p(l^-) = n, p(l^+)$, where $l = t_0 - W_{delption}(V)$ and $l = W_{delption}(V)$ for p-p-n and n-p-p, respectively.

Following the same procedures in Section 1.1, we can derive the equations for dark and photo currents ($V < V_{bi}$) following:

**Type 2 (p-p-n):**

$$\alpha_{f,ppn} = 1/(\Delta + \beta_f), \tag{S1.21}$$

$$\alpha_{b,ppn} = 1/(\Delta \times e^{V'} + \beta_b), \tag{S1.22}$$

$$A_{ppn} = \alpha_f \times \left(\frac{1}{m}(e^{-m\times\Delta} - 1) - \beta_f\right), \tag{S1.23}$$

$$B_{ppn} = \alpha_b \times \left(\frac{e^{V'}}{m}(e^{-m\times(\Delta-1)} - e^m) - \beta_b\right), \tag{S1.24}$$

**Type 4 (n-p-p):**

$$\alpha_{f,npp} = 1/(\Delta \times e^{V'} + \beta_f), \tag{S1.25}$$

$$\alpha_{b,npp} = 1/(\Delta + \beta_b), \tag{S1.26}$$

$$A_{npp} = \alpha_f \times \left(\frac{e^{V'}}{m}\left(e^{-m} - e^{m\times(\Delta-1)}\right) - \beta_f\right), \tag{S1.27}$$

$$B_{npp} = \alpha_b \times \left(\frac{1}{m}(1 - e^{m\times\Delta}) - \beta_b\right). \tag{S1.28}$$

The new parameter $\Delta = 1 - n\sqrt{(V_{bi} - V)/V_{bi}}$, where $n = W_{depletion}(0\,V)/t_0$ is the ratio of the equilibrium depletion width and the absorber thickness.

We assume that the self-doped absorber behaves identically as an intrinsic cell when $V \geq V_{bi}$. Hence we use Eqs. (S1.16) to (S1.20) to describe the operation of a self-doped device at $V \geq V_{bi}$. Please note that Eqs. (S1.16) to (S1.20) give the same limit as Eqs. (S1.21) to (S1.28) when $V \to V_{bi}$.

## 2 Fitting algorithm

The parameters of the compact model are extracted by fitting the equations to experimental data. The fitting algorithm has two parts: 1) Model choice 2) Iterative fitting. In the appendix, we demonstrate an illustrative MATLAB® script that can be used for fitting.

**2.1 Model choice**

Before one fits the data, the structure of the cell must be known (e.g., PEDOT: PSS/ Perovskite/PCBM or TiO$_2$/Perovskite/Spiro-OMeTAD) and whether the absorber is self-doped or not. Ideally, the capacitance-voltage measurement provides the doping profile; as an alternative, we find that the steepness (dI/dV) of the light I-V curve at low voltage can also differentiate self-doped and intrinsic cells, see Fig. S2.1. Specifically, the light IV of the self-doped device (sample #2) shows a steep decrease (~ 0 V – 0.5 V) in photocurrent much before the maximum power point (MPP); an undoped device (sample #1), however, shows flat light IV before MPP . If the parasitic resistance extracted from dark IV is not significant, our model attributes this decrease in photocurrent to voltage-dependent reduction of the depletion region (charge collection) of a doped absorber. Such a feature helps one to choose the correct model for a device.

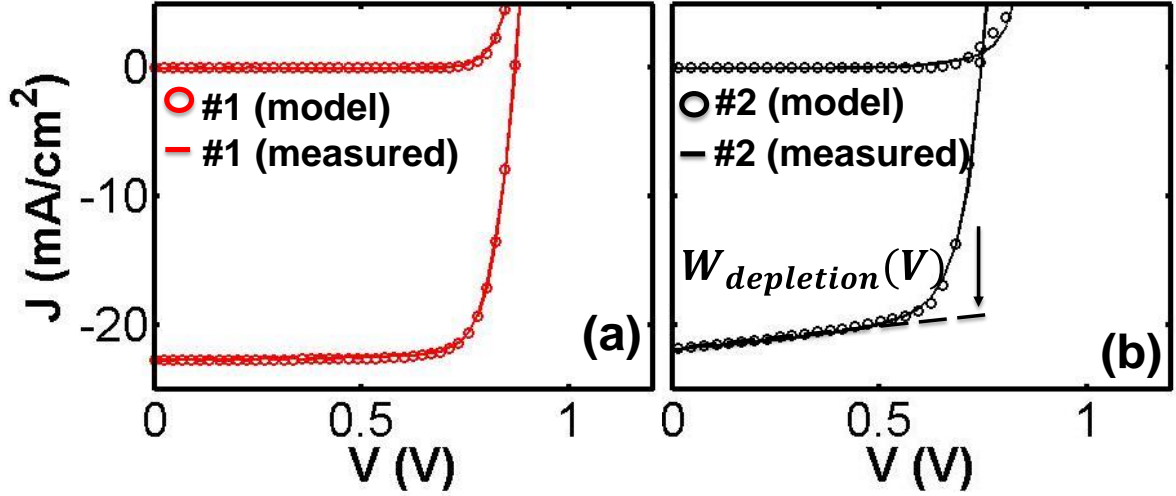

Figure S2.1 Fitting results of (a) Samples #1 (p-i-n, Efficiency = 15.7%, $J_{SC}$ = 22.7 mA/cm², $V_{OC}$ = 0.85 V, FF = 81%). (b) Samples #2 (p-p-n, Efficiency = 11.1%, $J_{SC}$ = 21.9 mA/cm², $V_{OC}$ = 0.75 V, FF = 64%).

## 2.1 Iterative fitting

Estimating the initial guesses and limiting the range of each parameter (from physical considerations) is an important step, since the fitting procedure utilize the iterative fitting function "lsqcurvefit" in MATLAB®, whose results depend on the initial guesses significantly.

The physical parameters we attempt to deduce are: $G_{max}$, $\lambda_{ave}$, $t_o$, $W_{depletion}(0\ V)$ (self-doped), $D$, $s_f$, $s_b$, $V_{bi}$, $J_{f0}$, and $J_{b0}$. Among these parameters, based on the transfer matrix method [3], $qG_{max}$ can be obtained by integrating the photon absorption (around 23 mA/cm²) and $\lambda_{ave}$ is around 100 nm; $D \approx 0.05\ cm^2s^{-1}$ is known for the material system for both electrons and holes.

### 2.1.1 Photocurrent

**Extracted physical parameter list: $t_o$, $W_{depletion}(0\ V)$ (self-doped), $s_f$, $s_b$, $V_{bi}$**

Presuming the dark current is illumination-independent, one can calculate photocurrent following

$$J_{photo}(G,V) = J_{light}(G,V) - J_{dark}(V). \tag{S2.1}$$

400 nm is a sensible initial guess for $t_o$, since the absorber thickness is around 300 nm to 500 nm for perovskite solar cells. Though capacitance measurement can determine $W_{depletion}(0\ V)$ for a self-doped device, one can make $W_{depletion}(0\ V) \approx 300$ nm as an initial guess. It has been shown that $s_f$ is inferior to $s_b$ in most cases due to low insufficient barrier between PEDOT:PSS and perovskites as well as low carrier lifetime in TiO₂. Hence, the initial guesses for $s_f$ and $s_b$ could be approximately $10^3$ cm/s and $10^2$ cm/s, respectively. The junction built-in $V_{bi}$ is estimated to be the cross-over voltage of dark and light IV curves.

Then one can use the "lsqcurvefit" function to fit the photocurrent based on the initial guesses.

### 2.1.2 Dark current

**Extracted physical parameter: $J_{f0}, J_{b0}$**

Since $J_{f0}$ and $J_{b0}$ is on the order of $10^{-13}$ to $10^{-15}$ mA/cm², one can use zero as the initial guesses. Afterwards, one can use the iterative fitting procedure for the dark current while the parameters extracted from photocurrent are fixed.

Once the parameters are obtained, they must be checked for self-consistency and convergence between light and dark characteristics.

# Appendix: Example Matlab script

```
function [coeff_final] = perovskite_fitting(JV)

% JV data format
%1st column is voltage (V)
%2nd column is light current (mA/cm2)
%3rd column is dark  current (mA/cm2)

% the list of the physical parameters
qgmax  = 23;      %mA/cm2
lambda = 100;     %nm
Dnp    = 0.05;    %0.05 cm2s-1
type = 3; % 1 for p-i-n/n-i-p; 2 for p-p-n; 3 for n-p-p;
global parms
parms =[qgmax;lambda;Dnp;type]; % set of input parameters
%vbi = coeff(1); %V
%to = coeff(2);   %nm
%sf = coeff(3);   %cm/s
%sb = coeff(4);   %cm/s
%jfo = coeff(5); %mA/cm2
%jbo = coeff(6); %mA/cm2
%wdepltion = coeff(7);   %nm

%calculate photocurrent
JPdataH=JV(:,2)-JV(:,3);
VdataH=JV(:,1);

%initial guess
coeff_init = [0.8;400;1e3;1e2;0;0; 300];

%fit photocurrent
% now we run optimization.
options = optimset('Display','iter','TolFun',1e-10,'TolX',1e-25);
% Constraints
lb=[0;     0;  1e-3; 1e-3;  0;  0; 0]; % lower  bound constraints
ub=[1.6;  500; 1e7;   1e7;   1;  1; 500]; % upper  bound constraints

[coeff_final,resnorm,residual,exitflag] =
lsqcurvefit(@pero_p,coeff_init,VdataH,JPdataH,lb,ub,options);

%plot photocurrent
figure(1)
plot(VdataH(:,1),pero_p(coeff_final,VdataH(:,1)),'or','LineWidth',2);
hold on
plot(VdataH(:,1),JPdataH,'-r','LineWidth',2);
set(gca,'LineWidth',2,'FontSize',22,'FontWeight','normal','FontName','Times')
set(get(gca,'XLabel'),'String','V
(V)','FontSize',22,'FontWeight','bold','FontName','Times')
set(get(gca,'YLabel'),'String','J
(mA/cm^2)','FontSize',22,'FontWeight','bold','FontName','Times')
set(gca,'box','on');
```

```matlab
%fit dark IV
coeffp = coeff_final;
pero_d2 = @(coeff,vd) pero_d(coeff,coeffp,vd);
lb=[0;      0; 1e-3; 1e-3;  0;  0; 0]; % lower  bound constraints
ub=[1.6;  500; 1e7;  1e7;  10; 10; 500]; % upper  bound constraints
[coeff_final,resnorm,residual,exitflag] =
lsqcurvefit(pero_d2,coeff_final,VdataH,JV(:,3),lb,ub,options);

%plot darkcurrent
figure(2)
plot(VdataH(:,1),pero_d2(coeff_final,VdataH(:,1)),'or','LineWidth',2);
hold on
plot(VdataH(:,1),JV(:,3),'r','LineWidth',2);
set(gca,'LineWidth',2,'FontSize',22,'FontWeight','normal','FontName','Times')
set(get(gca,'XLabel'),'String','V
(V)','FontSize',22,'FontWeight','bold','FontName','Times')
set(get(gca,'YLabel'),'String','J
(mA/cm^2)','FontSize',22,'FontWeight','bold','FontName','Times')
set(gca,'box','on');

coeff_final(5) = coeff_final(5)/1e10; %jfo normalized to mA/cm2
coeff_final(6) = coeff_final(6)/1e10; %jbo normalized to mA/cm2

%%function to calculate photocurrent
function [jphoto] = pero_p(coeff,vd)

qgmax  = parms(1);

lambda = parms(2);

Dnp    = parms(3);

type   = parms(4);

kt = 0.0259;

vbi = coeff(1)+1e-6; %for convergence
to = coeff(2);
sf = coeff(3);
sb = coeff(4);
wdelp = coeff(7);

m = to/lambda;
n = wdelp/to;
bf = Dnp/to/1e-7/sf;
bb = Dnp/to/1e-7/sb;
y = (vd-vbi)./kt;

if type == 1 % for p-i-n/n-i-p

    alphaf = 1./((exp(y)-1)./y+bf);
```

```matlab
    alphab = 1./((exp(y)-1)./y+bb);
    B = alphab .* ((1-exp(y+m))./(y+m)-bb);
    A = alphaf .* ((1-exp(y-m))./(y-m)-bf);

    jphoto = qgmax * (-B.*exp(-m)+A);

elseif type == 2 % for p-p-n

    yyy = 1 - n.* sqrt((vbi-vd)./vbi);

    for i = 1:length(vd)

        if vd(i) >= vbi

            alphaf = 1/((exp(y(i)))-1)/y(i)+bf);

            alphab = 1/((exp(y(i)))-1)/y(i)+bb);

            B = alphab * ((1-exp(y(i)+m))/(y(i)+m)-bb);

            A = alphaf * ((1-exp(y(i)-m))/(y(i)-m)-bf);

            jphoto(i) = qgmax   * (-B*exp(-m)+A);

        elseif vd(i) < vbi

            alphab = 1/(exp(y(i))*yyy(i)+bb);

            alphaf = 1/(yyy(i)+bf);

            A = alphaf * ((-1+exp(-yyy(i)*m))/m-bf);

            B = alphab * (exp(y(i))*(-exp(m)+exp(-m*(yyy(i)-1)))/m-bb);

            jphoto(i) = qgmax   * (-B*exp(-m)+A);

        end

    end

        jphoto = jphoto';

elseif type == 3 % for n-p-p

    yyy = 1 - n.* sqrt((vbi-vd)./vbi);

    for i = 1:length(vd)
```

```matlab
    if vd(i) >= vbi

        alphaf = 1/((exp(y(i))-1)/y(i)+bf);

        alphab = 1/((exp(y(i))-1)/y(i)+bb);

        B = alphab * ((1-exp(y(i)+m))/(y(i)+m)-bb);

        A = alphaf * ((1-exp(y(i)-m))/(y(i)-m)-bf);

        jphoto(i) = qgmax  * (-B*exp(-m)+A);

    elseif vd(i) < vbi

        alphaf = 1/(exp(y(i))*yyy(i)+bf);

        alphab = 1/(yyy(i)+bb);

        B = alphab * (-bb + (-exp(yyy(i)*m)+1)/m);

        A = alphaf * (exp(y(i))*(exp(-m)-exp(m*(yyy(i)-1)))/m-bf);

        jphoto(i) = qgmax  * (-B*exp(-m)+A);

     end

    end

        jphoto = jphoto';

end

end

%%function to calculate darkcurrent
function [jdark] = pero_d(coeff,coeffp,vd)

Dnp    = parms(3);
type   = parms(4);

kt = 0.0259;
vbi =coeffp(1)+1e-6; %for convergence;
to = coeffp(2);
sf = coeffp(3);
sb = coeffp(4);
jfo = coeff(5);
jbo = coeff(6);
```

```
wdelp = coeffp(7);
n = wdelp/to;
bf = Dnp/to/1e-7/sf;
bb = Dnp/to/1e-7/sb;
y = (vd-vbi)./kt;

if type == 1

    alphaf = 1./((exp(y)-1)./y+bf);
    alphab = 1./((exp(y)-1)./y+bb);
    %1e10 here just make it easy to converge
    jdark = (exp(vd/kt)-1).*(alphaf*jfo+alphab*jbo)/1e10;

elseif type == 2

         yyy = 1 - n.* sqrt((vbi-vd)./vbi);

    for i = 1:length(vd)

        if vd(i) < vbi

            alphab = 1/(exp(y(i))*yyy(i)+bb);

            alphaf = 1/(yyy(i)+bf);

            jdark(i) = (exp(vd(i)/kt)-1).*(alphaf*jfo+alphab*jbo)/1e10;

        else

            alphaf = 1./((exp(y(i))-1)./y(i)+bf);

            alphab = 1./((exp(y(i))-1)./y(i)+bb);

            jdark(i) = (exp(vd(i)/kt)-1).*(alphaf*jfo+alphab*jbo)/1e10;

        end

    end

        jdark =  jdark';

elseif type == 3

        yyy = 1 - n.* sqrt((vbi-vd)./vbi);

    for i = 1:length(vd)

        if vd(i) < vbi
```

```
            alphaf = 1/(exp(y(i))*yyy(i)+bf);

            alphab = 1/(yyy(i)+bb);

            jdark(i) = (exp(vd(i)/kt)-1)*(alphaf*jfo+alphab*jbo)/1e10;

        else

            alphaf = 1/((exp(y(i))-1)/y(i)+bf);

            alphab = 1/((exp(y(i))-1)/y(i)+bb);

            jdark(i) = (exp(vd(i)/kt)-1)*(alphaf*jfo+alphab*jbo)/1e10;

        end

    end

    jdark =  jdark';

end

end

end
```